Chapter 5

*Singular value decomposition and principal component analysis*
In A Practical Approach to Microarray Data Analysis *(D.P. Berrar, W. Dubitzky, M. Granzow, eds.) Kluwer: Norwell, MA, 2003. pp. 91-109. LANL LA-UR-02-4001*

Michael E. Wall[1,2], Andreas Rechtsteiner[1,3], Luis M. Rocha[1],
[1]*Computer and Computational Sciences Division and* [2]*Bioscience Division, Mail Stop B256, Los Alamos National Laboratory, Los Alamos, New Mexico, 87545 USA,* e-mail: {mewall, rocha}@lanl.gov
[3]*Systems Science Ph.D. Program, Portland State University, Post Office Box 751, Portland, Oregon 97207 USA,* e-mail: andreas@sysc.pdx.edu

Abstract: This chapter describes gene expression analysis by Singular Value Decomposition (SVD), emphasizing initial characterization of the data. We describe SVD methods for visualization of gene expression data, representation of the data using a smaller number of variables, and detection of patterns in noisy gene expression data. In addition, we describe the precise relation between SVD analysis and Principal Component Analysis (PCA) when PCA is calculated using the covariance matrix, enabling our descriptions to apply equally well to either method. Our aim is to provide definitions, interpretations, examples, and references that will serve as resources for understanding and extending the application of SVD and PCA to gene expression analysis.

## 1. INTRODUCTION

One of the challenges of bioinformatics is to develop effective ways to analyze global gene expression data. A rigorous approach to gene expression analysis must involve an up-front characterization of the structure of the data. In addition to a broader utility in analysis methods, singular value decomposition (SVD) and principal component analysis (PCA) can be valuable tools in obtaining such a characterization. SVD and PCA are common techniques for analysis of multivariate data, and gene expression data are well suited to analysis using SVD/PCA. A single *microarray*[1] experiment can generate measurements for thousands, or even tens of thousands of genes. Present experiments typically consist of less than ten assays, but can consist of hundreds (Hughes *et al.*, 2000). Gene expression data are currently rather noisy, and SVD can detect and extract small signals from noisy data.

The goal of this chapter is to provide precise explanations of the use of SVD and PCA for gene expression analysis, illustrating methods using simple examples. We describe SVD methods for visualization of gene expression data, representation of the data using a smaller number of variables, and detection of patterns in noisy gene expression data. In addition, we describe the mathematical relation between SVD analysis and Principal Component Analysis (PCA) when PCA is calculated using the covariance matrix, enabling our descriptions to apply equally well to either method. Our aims are 1) to provide descriptions and examples of the application of SVD methods and interpretation of their results; 2) to establish a foundation for understanding previous applications of SVD to gene expression analysis; and 3) to provide interpretations and references to related work that may inspire new advances.

In section 1, the SVD is defined, with associations to other methods described. A summary of previous applications is presented in order to suggest directions for SVD analysis of gene expression data. In section 2 we discuss applications of SVD to gene expression analysis, including specific methods for SVD-based visualization of gene expression data, and use of SVD in detection of weak expression patterns. Some examples are given of previous applications of SVD to analysis of gene expression data. Our discussion in section 3 gives some general advice on the use of SVD analysis on gene expression data, and includes references to specific published SVD-based methods for gene expression analysis. Finally, in section 4, we provide information on some available resources and further reading.

## 1.1    Mathematical definition of the SVD[2]

Let $X$ denote an $m \times n$ matrix of real-valued data and *rank*[3] $r$, where without loss of generality $m \geq n$, and therefore $r \leq n$. In the case of microarray data, $x_{ij}$ is the expression level of the $i^{th}$ gene in the $j^{th}$ assay. The elements of the $i^{th}$ row of $X$ form the $n$-dimensional vector $\mathbf{g}_i$, which we refer to as the *transcriptional response* of the $i^{th}$ gene. Alternatively, the elements of the $j^{th}$ column of $X$ form the $m$-dimensional vector $\mathbf{a}_j$, which we refer to as the *expression profile* of the $j^{th}$ assay.

The equation for singular value decomposition of $X$ is the following:

$$X = USV^{T}, \qquad (5.1)$$

where $U$ is an $m \times n$ matrix, $S$ is an $n \times n$ diagonal matrix, and $V^T$ is also an $n \times n$ matrix. The columns of $U$ are called the *left singular vectors*, $\{\mathbf{u}_k\}$, and form an orthonormal basis for the assay expression profiles, so that $\mathbf{u}_i \cdot \mathbf{u}_j = 1$ for $i = j$, and $\mathbf{u}_i \cdot \mathbf{u}_j = 0$ otherwise. The rows of $V^T$ contain the elements of the *right singular vectors*, $\{\mathbf{v}_k\}$, and form an orthonormal basis for the gene transcriptional responses. The elements of $S$ are only nonzero on the diagonal, and are called the *singular values*. Thus, $S = \text{diag}(s_1,...,s_n)$. Furthermore, $s_k > 0$ for $1 \leq k \leq r$, and $s_i = 0$ for $(r+1) \leq k \leq n$. By convention, the ordering of the singular vectors is determined by high-to-low sorting of singular values, with the highest singular value in the upper left index of the $S$ matrix. Note that for a square, symmetric matrix $X$, singular value decomposition is equivalent to diagonalization, or solution of the eigenvalue problem.

One important result of the SVD of $X$ is that

$$X^{(l)} = \sum_{k=1}^{l} \mathbf{u}_k s_k \mathbf{v}_k^{T} \qquad (5.2)$$

is the closest rank-$l$ matrix to $X$. The term "closest" means that $X^{(l)}$ minimizes the sum of the squares of the difference of the elements of $X$ and $X^{(l)}$, $\sum_{ij}|x_{ij} - x^{(l)}_{ij}|^2$.

One way to calculate the SVD is to first calculate $V^T$ and $S$ by diagonalizing $X^T X$:

$$X^TX = VS^2V^T, \quad (5.3)$$

and then to calculate $U$ as follows:

$$U = XVS^{-1}, \quad (5.4)$$

where the $(r+1),...,n$ columns of $V$ for which $s_k = 0$ are ignored in the matrix multiplication of Equation 5.4. Choices for the remaining $n-r$ singular vectors in $V$ or $U$ may be calculated using the Gram-Schmidt orthogonalization process or some other extension method. In practice there are several methods for calculating the SVD that are of higher accuracy and speed. Section 4 lists some references on the mathematics and computation of SVD.

*Relation to principal component analysis*. There is a direct relation between PCA and SVD in the case where principal components are calculated from the *covariance matrix*[4]. If one conditions the data matrix $X$ by *centering*[5] each column, then $X^TX = \Sigma_i \mathbf{g}_i \mathbf{g}_i^T$ is proportional to the covariance matrix of the variables of $\mathbf{g}_i$ (*i.e.*, the covariance matrix of the assays[6]). By Equation 5.3, diagonalization of $X^TX$ yields $V^T$, which also yields the principal components of $\{\mathbf{g}_i\}$. So, the right singular vectors $\{\mathbf{v}_k\}$ are the same as the principal components of $\{\mathbf{g}_i\}$. The eigenvalues of $X^TX$ are equivalent to $s_k^2$, which are proportional to the variances of the principal components. The matrix $US$ then contains the *principal component scores*, which are the coordinates of the genes in the space of principal components.

If instead each row of $X$ is centered, $XX^T = \Sigma_j \mathbf{a}_j \mathbf{a}_j^T$ is proportional to the covariance matrix of the variables of $\mathbf{a}_j$ (*i.e.* the covariance matrix of the genes[7]). In this case, the left singular vectors $\{\mathbf{u}_k\}$ are the same as the principal components of $\{\mathbf{a}_j\}$. The $s_k^2$ are again proportional to the variances of the principal components. The matrix $SV^T$ again contains the principal component scores, which are the coordinates of the assays in the space of principal components.

*Relation to Fourier analysis*. Application of SVD in data analysis has similarities to Fourier analysis. As is the case with SVD, Fourier analysis involves expansion of the original data in an orthogonal basis:

$$x_{ij} = \sum_k c_{ik} e^{i2\pi jk/m} \quad (5.5)$$

The connection with SVD can be explicitly illustrated by *normalizing*[8] the vector $\{e^{i2\pi jk/m}\}$ and by naming it $\mathbf{v'}_k$:

$$x_{ij} = \sum_k b_{ik} v'_{jk} = \sum_k u'_{ik} s'_k v'_{jk} \quad (5.6)$$

which generates the matrix equation $X = U'S'V'^T$, similar to Equation 5.1. Unlike the SVD, however, even though the $\{\mathbf{v'}_k\}$ are an orthonormal basis, the $\{\mathbf{u'}_k\}$ are not in general orthogonal. Nevertheless this demonstrates how the SVD is similar to a Fourier transform, where the vectors $\{\mathbf{v}_k\}$ are determined in a very specific way from the data using Equation 5.1, rather than being given at the outset as for the Fourier transform. Similar to low-pass filtering in Fourier analysis, later we will describe how SVD analysis permits filtering by concentrating on those singular vectors that have the highest singular values.

## 1.2    Illustrative applications

SVD and PCA have found wide-ranging applications. Here we describe several that may suggest potential ways that we can think about applications in gene expression analysis.

*Image processing and compression*. The property of SVD to provide the closest rank-$l$ approximation for a matrix $X$ (Equation 5.2) can be used in image processing for compression and noise reduction, a very common application of SVD. By setting the small singular values to zero, we can obtain matrix approximations whose rank equals the number of remaining singular values (see Equation 5.2). Each term $\mathbf{u}_k s_k \mathbf{v}_k^T$ is called a *principal image*. Very good approximations can often be obtained using only a small number of terms (Richards, 1993). SVD is applied in similar ways to signal processing problems (Deprettere, 1988).

*Immunology*. One way to capture global prototypical immune response patterns is to use PCA on data obtained from measuring antigen-specific IgM (dominant antibody in primary immune responses) and IgC (dominant antibody in secondary immune responses) immunoglobulins using ELISA assays. Fesel and Coutinho (Fesel and Coutinho, 1998) measured IgM and IgC responses in Lewis and Fischer rats before and at three time points after immunization with myelin basic protein (MBP) in complete Freud's adjuvant (CFA), which is known to provoke experimental allergic encephalomeyelitis (EAE). They discovered distinct and mutually independent components of IgM reaction repertoires, and identified a small number of strain-specific prototypical regulatory responses.

*Molecular dynamics*. PCA and SVD analysis methods have been developed for characterizing protein molecular dynamics trajectories (Garcia, 1992; Romo *et al.*, 1995). In a study of myoglobin, Romo *et al.* used molecular dynamics methods to obtain atomic positions of all atoms sampled during the course of a simulation. The higher principal components of the dynamics were found to correspond to large-scale motions of the protein. Visualization of the first three principal components revealed an interesting type of trajectory that was described as resembling beads on a string, and revealed a visibly sparse sampling of the configuration space.

*Small-angle scattering*. SVD has been used to detect and characterize structural intermediates in biomolecular small-angle scattering experiments (Chen *et al.*, 1996). This study provides a good illustration of how SVD can be used to extract biologically meaningful signals from the data. Small-angle scattering data were obtained from partially unfolded solutions of lysozyme, each consisting of a different mix of folded, collapsed and unfolded states. The data for each sample was in the form of intensity values sampled at on the order of 100 different scattering angles. UV spectroscopy was used to determine the relative amounts of folded, collapsed and unfolded lysozyme in each sample. SVD was used in combination with the spectroscopic data to extract a scattering curve for the collapsed state of the lysozyme, a structural intermediate that was not observed in isolation.

*Information Retrieval*. SVD became very useful in Information Retrieval (IR) to deal with linguistic ambiguity issues. IR works by producing the documents most associated with a set of keywords in a query. Keywords, however, necessarily contain much synonymy (several keywords refer to the same concept) and polysemy (the same keyword can refer to several concepts). For instance, if the query keyword is "feline", traditional IR methods will not retrieve documents using the word "cat" – a problem of synonymy. Likewise, if the query keyword is "java", documents on the topic of Java as a computer language, Java as an Island in Indonesia, and Java as a coffee bean will all be retrieved – a problem of polysemy. A technique known *Latent Semantic Indexing* (LSI) (Berry *et al.*, 1995) addresses these problems by calculating the best rank-$l$ approximation of the keyword-document matrix using its SVD. This produces a lower dimensional space of singular vectors that are called *eigen-keywords* and *eigen-documents*. Each eigen-keyword can be associated with several keywords as well as particular senses of keywords. In the synonymy example above, "cat" and "feline" would therefore be strongly correlated with the same eigen-keyterm. Similarly, documents using "Java" as a computer language tend to use many of the same keywords, but not many of the keywords used by documents describing "Java" as coffee or Indonesia. Thus, in the space of singular vectors, each of these senses of "java" is associated with distinct eigen-keywords.

## 2. SVD ANALYSIS OF GENE EXPRESSION DATA

As we mention in the introduction, gene expression data are well suited to analysis using SVD/PCA. In this section we provide examples of SVD-based analysis methods as applied to gene expression analysis. Before illustrating specific techniques, we will discuss ways of interpreting the SVD in the context of gene expression data. This interpretation and the accompanying nomenclature will serve as a foundation for understanding the methods described later.

A natural question for a biologist to ask is: "What is the biological significance of the SVD?" There is, of course, no general answer to this question, as it depends on the specific application. We can, however, consider classes of experiments and provide them as a guide for individual cases. For this purpose we define two broad classes of applications under which most studies will fall: *systems biology applications*, and *diagnostic applications* (see below). In both cases, the $n$ columns of the gene expression data matrix $X$ correspond to assays, and the $m$ rows correspond to the genes. The SVD of $X$ produces two orthonormal bases, one defined by right singular vectors and the other by left singular vectors. Referring to the definitions in section 1.1, the right singular vectors span the space of the gene transcriptional responses $\{\mathbf{g}_i\}$ and the left singular vectors span the space of the assay expression profiles $\{\mathbf{a}_j\}$. Following the convention of (Alter *et al.*, 2000), we refer to the left singular vectors $\{\mathbf{u}_k\}$ as *eigenassays* and to the right singular vectors $\{\mathbf{v}_k\}$ as *eigengenes*[9]. We sometimes refer to an eigengene or eigenassay generically as a singular vector, or, by analogy with PCA, as a *component*. Eigengenes, eigenassays and other definitions and nomenclature in this section are depicted in Figure 5.1.

In systems biology applications, we generally wish to understand relations among genes. The signal of interest in this case is the gene transcriptional response $\mathbf{g}_i$. By Equation 5.1, the SVD equation for $\mathbf{g}_i$ is

$$\mathbf{g}_i = \sum_{k=1}^{r} u_{ik} s_k \mathbf{v}_k, \quad i:1,...,m \qquad (5.7)$$

which is a linear combination of the eigengenes $\{\mathbf{v}_k\}$. The $i^{th}$ row of $U$, $\mathbf{g}'_i$ (see Figure 5.1), contains the coordinates of the $i^{th}$ gene in the coordinate system (basis) of the scaled eigengenes, $s_k \mathbf{v}_k$. If $r < n$, the transcriptional responses of the genes may be captured with fewer variables using $\mathbf{g}'_i$ rather than $\mathbf{g}_i$. This property of the SVD is sometimes referred to as *dimensionality reduction*. In order to reconstruct the original data, however, we still need access to the eigengenes, which are $n$-dimensional vectors. Note that due to the presence of noise in the measurements, $r = n$ in any real gene expression analysis application, though the last singular values in $S$ may be very close to zero and thus irrelevant.

In diagnostic applications, we may wish to classify tissue samples from individuals with and without a disease. Referring to the definitions in section 1.1, the signal of interest in this case is the assay expression profile $\mathbf{a}_j$. By Equation 5.1, the SVD equation for $\mathbf{a}_j$ is

$$\mathbf{a}_j = \sum_{k=1}^{r} v_{jk} s_k \mathbf{u}_k, \quad j:1,\ldots,n \qquad (5.8)$$

which is a linear combination of the eigenassays $\{\mathbf{u}_k\}$. The $j^{\text{th}}$ column of $V^T$, $\mathbf{a}'_j$ (see Figure 5.1), contains the coordinates of the $j^{\text{th}}$ assay in the coordinate system (basis) of the scaled eigenassays, $s_k \mathbf{u}_k$. By using the vector $\mathbf{a}'_j$, the expression profiles of the assays may be captured by $r \leq n$ variables, which is always fewer than the $m$ variables in the vector $\mathbf{a}_j$. So, in contrast to gene transcriptional responses, SVD can generally reduce the number of variables used to represent the assay expression profiles. Similar to the case for genes, however, in order to reconstruct the original data, we need access to the eigenassays, which are $m$-dimensional vectors.

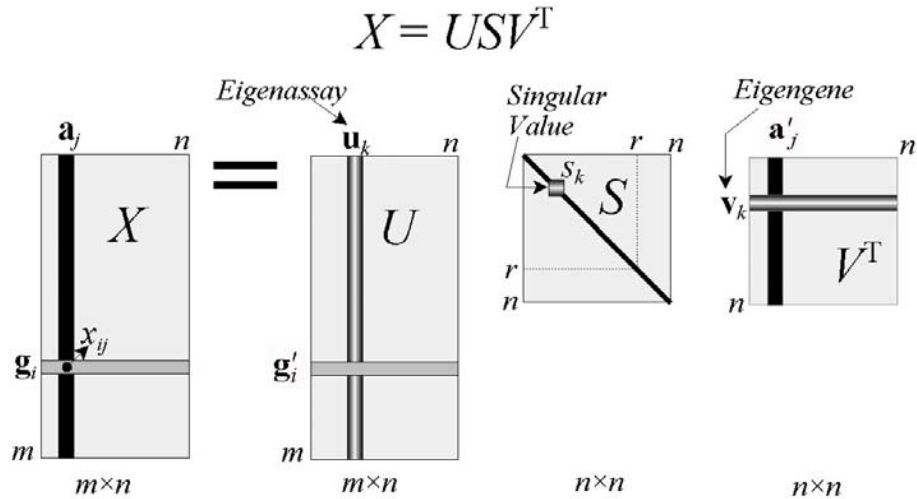

Figure 5.1. Graphical depiction of SVD of a matrix *X*, annotated with notations adopted in this chapter.

Indeed, analysis of the spectrum formed by the singular values $s_k$ can lead to the determination that fewer than $n$ components capture the essential features in the data, a topic discussed below in section 2.1.1. In the literature the number of components that results from such an analysis is sometimes associated with the number of underlying biological processes that give rise to the patterns in the data. It is then of interest to ascribe biological meaning to the significant eigenassays (in the case of diagnostic applications), or eigengenes (in the case of systems biology applications). Even though each component on its own may not necessarily be biologically meaningful, SVD can aid in the search for biologically meaningful signals (see, *e.g.*, small-angle scattering in section 1.2).

In the context of describing scatter plots in section 2.1.2, we discuss the application of SVD to the problem of grouping genes by transcriptional response, and grouping assays by expression profile. This discussion will also touch on the topic of searching for biologically meaningful signals. When the data are noisy, it may not be possible to resolve gene groups, but it still may be of interest to detect underlying gene expression patterns; this is a case where the utility of the SVD distinguishes itself with respect to other gene expression analysis methods (section 2.2). Finally we discuss some published examples of gene expression analysis using SVD, and a couple of SVD-based gene grouping methods (section 2.3).

## 2.1 Visualization of the SVD

Visualization is central to understanding the results of application of SVD to gene expression data. For example, Figure 5.2 illustrates plots that are derived from applying SVD to Cho *et al.*'s budding yeast cell-cycle data set (Cho *et al.*, 1998). In the experiment, roughly 6,200 yeast genes were monitored for 17 time points taken at ten-minute intervals. To perform the SVD, we have pre-processed the data by replacing each measurement with its logarithm, and normalizing each gene's transcriptional response to have zero mean and unit standard deviation. In addition, a serial correlation test (Kanji, 1993) was applied to filter out ~3,200 genes that showed primarily random fluctuations. The plots reveal interesting patterns in the data that we may wish to investigate further: a levelling off of the relative variance after the first five components (Figure 5.2a); a pattern in the first eigengene primarily resembling a steady decrease, or decay (Figure 5.2b); and patterns with cyclic structure in the second and third eigengenes (Figure 5.2c,d).

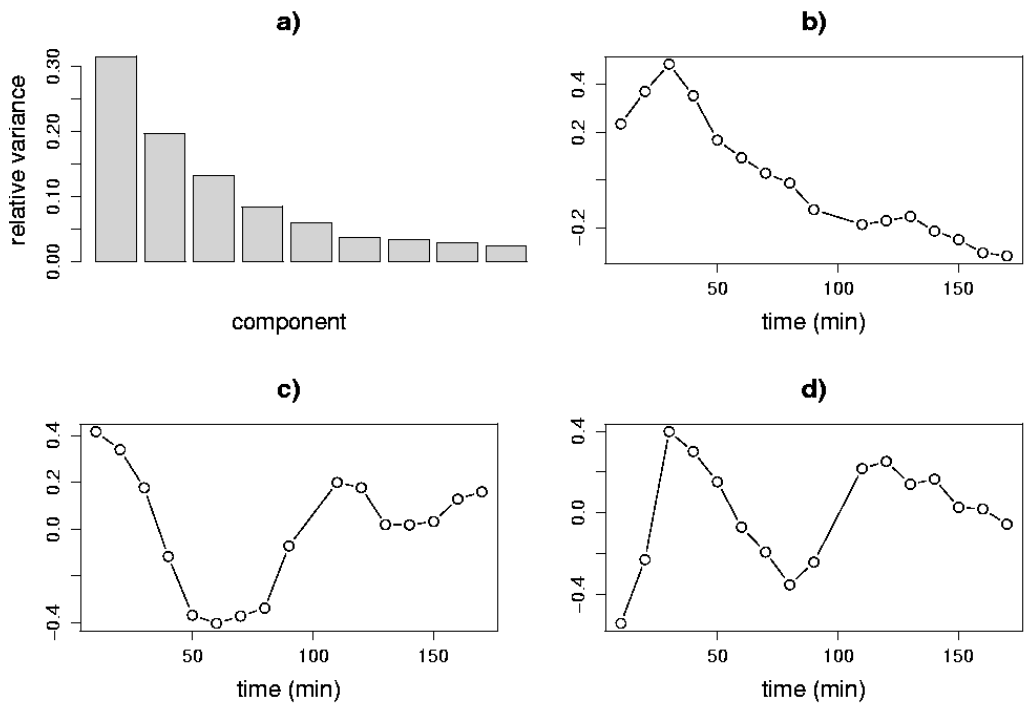

Figure 5.2. Visualization of the SVD of cell cycle data. Plots of relative variance (a); and the first (b), second (c) and third (d) eigengenes are shown. The methods of visualization employed in each panel are described in section 2.1. These data inspired our choice of the sine and exponential patterns for the synthetic data of section 2.1.

To aid our discussion of visualization, we use a synthetic time series data set with 14 sequential expression level assays (columns of $X$) of 2,000 genes (rows of $X$). Use of a synthetic data set enables us to provide simple illustrations that can serve as a foundation for understanding the more complex patterns that arise in real gene expression data. Genes in our data set have one of three kinds of transcriptional response, inspired by experimentally observed patterns in the Cho *et al.* cell-cycle data: 1) noise (1,600 genes); 2) noisy sine pattern (200 genes); or 3) noisy exponential pattern (200 genes). Noise for all three groups of genes was modelled by sampling from a normal distribution with zero mean and standard deviation 0.5. The sine pattern has the functional form $a\sin(2\pi t/140)$, and the exponential pattern the form $be^{-t/100}$, where $a$ is sampled uniformly over the interval (1.5,3), $b$ is sampled uniformly over (4,8), $t$ is the time (in minutes) associated with each assay, and time points are sampled every ten minutes beginning at $t = 0$. Each gene's transcriptional response was centered to have a mean of zero. Figure 5.3 depicts genes of type 2) and 3).

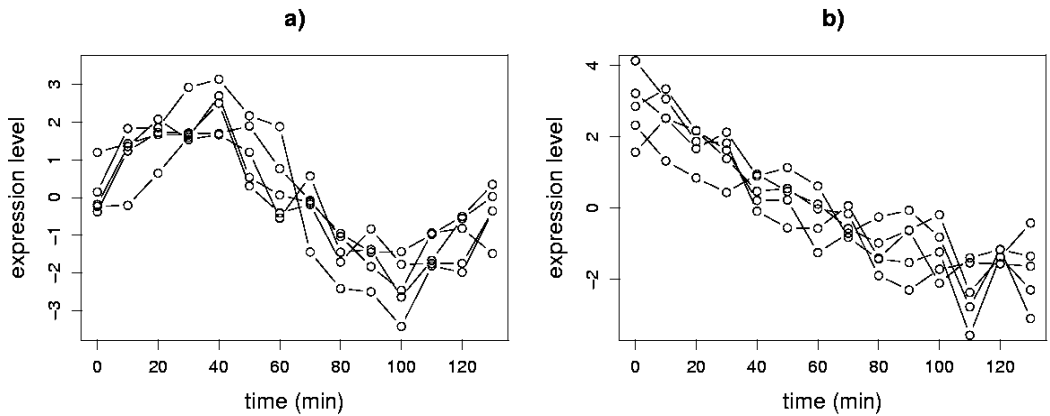

Figure 5.3. Gene transcriptional responses from the synthetic data set. Overlays of a) five noisy sine wave genes and b) five noisy exponential genes.

### 2.1.1    Visualization of the matrices *S*, $V^T$ and *U*

*Singular value spectrum*. The diagonal values of *S* (*i.e.*, $s_k$) make up the singular value spectrum, which is easily visualized in a one-dimensional plot. The height of any one singular value is indicative of its importance in explaining the data. More specifically, the square of each singular value is proportional to the variance explained by each singular vector. The relative variances $s_k^2(\sum_i s_i^2)^{-1}$ are often plotted (Figure 5.4a; see also Figure 5.2). Cattell has referred to these kinds of plots as *scree plots* (Cattell, 1966) and proposed to use them as a graphical method to decide on the significant components. If the original variables are linear combinations of a smaller number of underlying variables, combined with some low-level noise, the plot will tend to drop sharply for the singular values associated with the underlying variables and then much more slowly for the remaining singular values. Singular vectors (in our case eigenassays and eigengenes) whose singular values plot to the right of such an "elbow" are ignored because they are assumed to be mainly due to noise. For our synthetic data set, the spectrum begins with a sharp decrease, and levels off after the second component, which is indicative of the two underlying signals in the data (Figure 5.4a). Other heuristic approaches for deciding on the significant components have been proposed. One approach is to ignore components beyond where the cumulative relative variance or singular value becomes larger than a certain threshold, usually defined upon the dimensionality of the data. For our example data set, the first two singular vectors explain about 64% of the total variance in the data (Figure 5.4a). Everitt and Dunn propose an alternate approach based on comparing the relative variance of each component to $0.7/n$ (Everitt and Dunn, 2001). For our example data set this threshold is $(0.7/14) = 0.05$, which selects the first two singular vectors as significant. Notice that if we re-construct the matrix $X$ by using only the first two singular vectors, we would obtain $X^{(2)}$ (the best rank-2 approximation of $X$), which would account for 64% of the variance in the data.

*Eigengenes.* When assays correspond to samplings of an ordinal or continuous variable (*e.g.*, time; radiation dose; toxin concentration), a plot of the elements of the eigengenes $\{\mathbf{v}_k\}$ may reveal recognizable patterns. In our example, the first two eigengenes show an obvious cyclic structure (Figure 5.4b,c; see also Figure 5.2). Neither eigengene is exactly like the underlying sine or exponential pattern; each such pattern, however, is closely approximated by a linear combination of the eigengenes. Sine wave and exponential patterns cannot simultaneously be right singular vectors, as they are not orthogonal. This illustrates the point that, although the most significant eigengenes may not be biologically meaningful in and of themselves, they may be linearly combined to form biologically meaningful signals.

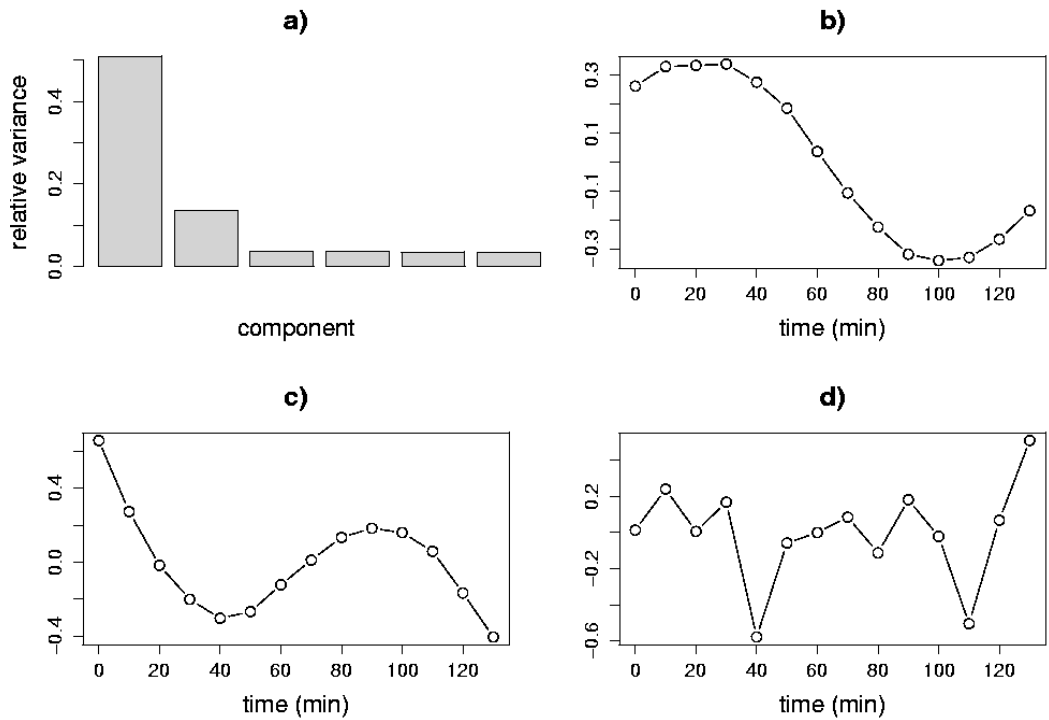

Figure 5.4. Visualization of the SVD of the synthetic data matrix. a) Singular value spectrum in a relative variance plot. The first two singular values account for 64% of the variance. The first (b), second (c), and third (d) eigengenes are plotted vs. time (assays) in the remaining panels. The third eigengene lacks the obvious cyclic structure of the first and second.

When assays correspond to discrete experimental conditions (*e.g.*, mutational varieties; tissue types; distinct individuals), visualization schemes are similar to those described below for eigenassays. When the $j^{th}$ element of eigengene $k$ is of large-magnitude, the $j^{th}$ assay is understood to contribute relatively strongly to the variance of eigenassay $k$, a property that may be used for associating a group of assays.

   *Eigenassays.* Alter *et al.* have visualized eigenassays {$\mathbf{u}_k$} resulting from SVD analysis of cell-cycle data (Alter *et al.*, 2000) by adapting a previously developed color-coding scheme for visualization of gene expression data matrices (Eisen *et al.*, 1998). Individual elements of $U$ are displayed as rectangular pixels in an image, and color-coded using green for negative values, and red for positive values, the intensity being correlated with the magnitude. The rows of matrix $U$ can be sorted using correlation to the eigengenes. In Alter *et al.*'s study, this scheme sorted the genes by the phase of their periodic pattern. The information communicated in such visualization bears some similarity to visualization using scatter plots, with the advantage that the table-like display enables gene labels to be displayed along with the eigenassays, and the disadvantage that differences among the genes can only be visualized in one dimension.

## 2.1.2    Scatter plots

   Visualization of structure in high-dimensional data requires display of the data in a one-, two-, or three-dimensional subspace. SVD identifies subspaces that capture most of the variance in the data. Even though our discussion here is about visualization in subspaces obtained by SVD, the illustrated visualization techniques are general and can in most cases be applied for visualization in other subspaces (see section 4 for techniques that use other criteria for subspace selection).

   For gene expression analysis applications, we may want to classify samples in a diagnostic study, or classify genes in a systems biology study. Projection of data into SVD subspaces and visualization with scatter plots can reveal structures in the data that may be used for classification. Here we discuss the visualization of features that may help to distinguish gene groups by transcriptional response. Analogous methods are used to distinguish groups of assays by expression profile. We discuss two different sources of gene "coordinates" for scatter plots: projections of the transcriptional response onto eigengenes, and correlations of the transcriptional response with eigengenes.

   *Projection and correlation scatter plots.* Projection scatter plot coordinates $q_{ik}$ for transcriptional response $\mathbf{g}_i$ projected on eigengene $\mathbf{v}_k$ are calculated as $q_{ik} = \mathbf{g}_i \cdot \mathbf{v}_k$. The SVD of $X$ readily allows computation of these coordinates using the equation $XV = US$, so that $q_{ik} = (US)_{ik}$. The projection of gene transcriptional responses from our example data onto the first two eigengenes reveals the *a priori* structure in the data (Figure 5.5a). The groups of the 200 sine wave genes (bottom right cluster), and the 200 exponential decay genes (top right cluster) are clearly separated from each other and from the 1,600 pure noise genes, which cluster about the origin.

   Correlation scatter plots may be obtained by calculating the Pearson correlation coefficient of each gene's transcriptional response with the eigengenes:

$$r_{ik} = \delta \mathbf{g}_i \cdot \delta \mathbf{v}_k \left| \delta \mathbf{g}_i \right|^{-1} \left| \delta \mathbf{v}_k \right|^{-1} \qquad (5.9)$$

where $r_{ik}$ denotes the correlation coefficient of the transcriptional response $\mathbf{g}_i$ with eigengene $\mathbf{v}_k$; $\delta\mathbf{g}_i$ is the mean-centered $\mathbf{g}_i$, the elements of which are $\{x_{ij} - \langle x_{ij}\rangle_j\}_i$, and $\delta\mathbf{v}_k$ is the mean-centered $\mathbf{v}_k$, the elements of which are $\{v_{jk} - \langle v_{jk}\rangle_j\}_k$. The normalization leads to $-1 \leq r_{ik} \leq 1$. Note that if each $\mathbf{g}_i$ is pre-processed to have zero mean and unit norm, it follows that the correlation scatter plot is equivalent to the projection scatter plot ($\mathbf{g}_i = \delta\mathbf{g}_i$ implies $\mathbf{v}_k = \delta\mathbf{v}_k$; and $|\delta\mathbf{g}_i|^{-1} = |\delta\mathbf{v}_k|^{-1} = 1$).

In the projection scatter plot, genes with a relatively high-magnitude coordinate on the *k*-axis contribute relatively strongly to the variance of the $k^{th}$ eigengene in the data set. The farther a gene lies away from the origin, the stronger the contribution of that gene is to the variance accounted for by the subspace. In the correlation scatter plot, genes with a relatively high-magnitude coordinate on the *k*-axis have transcriptional responses that are relatively highly correlated with the $k^{th}$ eigengene.

Due to the normalization in correlation scatter plots, genes with similar patterns in their transcriptional responses, but with different amplitudes, can appear to cluster more tightly in a correlation scatter plot than in a projection scatter plot. Genes that correlate well with the eigengenes lie near the perimeter, a property that can be used in algorithms that seek to identify interesting genes. At the same time, low-amplitude noise genes can appear to be magnified in a correlation scatter plot. For our example data, the sine wave and exponential gene clusters are relatively tightened, the scatter of the noise genes appears to be increased, and the separation between signal and noise genes is decreased for the correlation *vs.* the projection scatter plot (Figure 5.5).

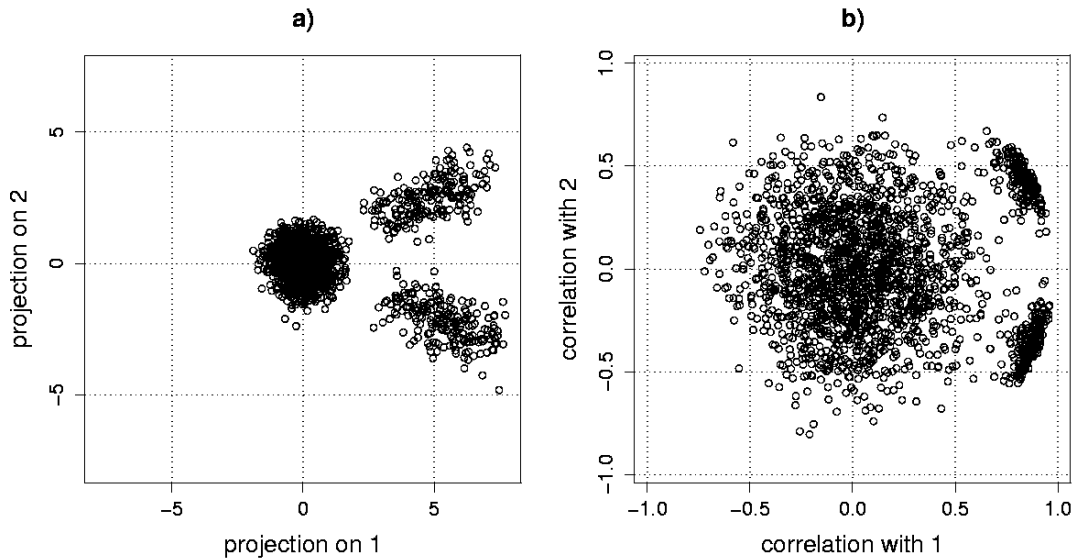

Figure 5.5. SVD scatter plots. Genes from our synthetic example data set are displayed in a) a projection scatter plot; and b) a correlation scatter plot. The bottom right cluster corresponds to sine wave genes, and the top right cluster corresponds to exponential decay genes. The cluster of genes around the origin corresponds to the noise-only genes.

The projection scatter plot (Figure 5.5a) illustrates how SVD may be used to aid in detection of biologically meaningful signals. In this case, the position ($q_1$, $q_2$) of any *cluster center*[10] may be used to construct the cluster's transcriptional response **g** from the right singular vectors:

$$\mathbf{g} = q_1 \mathbf{v}_1 + q_2 \mathbf{v}_2 \qquad (5.10)$$

If the first and second singular vectors are biologically meaningful in and of themselves, the cluster centers will lie directly on the axes of the plot. For our synthetic data, the first and second singular vectors are combined to approximately generate the sine wave and exponential patterns.

SVD and related methods are particularly valuable analysis methods when the distribution of genes is more complicated than the simple distributions in our example data: for instance, SVD has been used to characterize ring-like distributions of genes such as are observed in scatter plots of cell-cycle gene expression data (Alter *et al.*, 2000; Holter *et al.*, 2000) (see section 2.3).

*Scatter plots of assays*. Assays can be visualized in scatter plots using methods analogous to those used for genes. Coordinates for projection scatter plots are obtained by taking the dot products $\mathbf{a}_j \cdot \mathbf{u}_k$ of expression profiles on eigenassays, and coordinates for correlation scatter plots are obtained by calculating the Pearson correlation coefficient $\delta\mathbf{a}_j \cdot \delta\mathbf{u}_k |\delta\mathbf{a}_j|^{-1} |\delta\mathbf{u}_k|^{-1}$. Such plots are useful for visualizing diagnostic data, *e.g.*, distinguishing groups of individuals according to expression profiles. Alter *et al.* used such a technique to visualize cell-cycle assays (Alter *et al.*, 2000), and were able to associate individual assays with different phases of the cell cycle.

## 2.2 Detection of weak expression patterns

As noise levels in the data increase, it is increasingly difficult to obtain separation of gene groups in scatter plots. In such cases SVD may still be able to detect weak patterns in the data that may be associated with biological effects. In this respect SVD and related methods provide information that is unique among commonly used analysis methods.

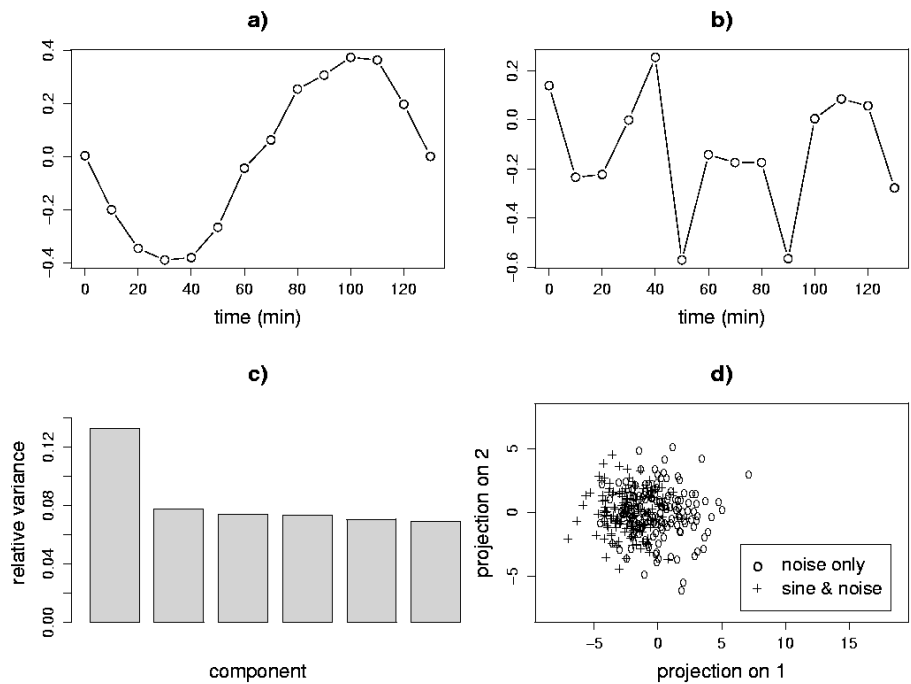

Figure 5.6. SVD-based detection of weak signals. a) A plot of the first eigengene shows the structure of the weak sine wave signal that contributes to the transcriptional response for half of the genes. b) The second eigengene resembles noise. c) A relative variance plot for the first six singular values shows an elbow after the first singular value. d) The signal and noise genes are not separated in an eigengene scatter plot of 150 of the signal genes, and 150 of the noise-only genes.

Here we will use an example to illustrate the ability of SVD to detect patterns in gene transcriptional response even though the individual genes may not clearly separate in a scatter plot. A data matrix was generated using two kinds of transcriptional response: 1,000 genes exhibiting a sine pattern, $\sin(2\pi t/140)$, with added noise sampled from a normal distribution of zero mean and standard deviation 1.5; and 1,000 genes with just noise sampled from the same distribution. Upon application of SVD, we find that the first eigengene shows a coherent sine pattern (Figure 5.6a). The second eigengene is dominated by high-frequency components that can only come from the noise (Figure 5.6b), and the singular value spectrum has an elbow after the first singular value (Figure 5.6c), suggesting (as we know *a priori*) that there is only one interesting signal in the data. Even though the SVD detected the cyclic pattern in the first eigengene (Figure 5.6a), the sine wave and noise-only genes are not clearly separated in the SVD eigengene projection scatter plot (Figure 5.6d).

## 2.3 Examples from the literature

Cell-cycle gene expression data display strikingly simple patterns when analyzed using SVD. Here we discuss two different studies that, despite having used different pre-processing methods, have produced similar results (Alter *et al.*, 2000; Holter *et al.*, 2000). Both studies found cyclic patterns for the first two eigengenes, and, in two-dimensional correlation scatter plots, previously identified cell cycle genes tended to plot towards the perimeter of a disc. Alter *et al.* used information in SVD correlation scatter plots to obtain a result that 641 of the 784 cell-cycle genes identified in (Spellman *et al.*, 1998) are associated with the first two eigengenes. Holter *et al.* displayed previously identified cell-cycle gene clusters in scatter plots, revealing that cell-cycle genes were relatively uniformly distributed in a ring-like feature around the perimeter, leading Holter *et al.* to suggest that cell-cycle gene regulation may be a more continuous process than had been implied by the previous application of clustering algorithms.

Raychaudhuri *et al.*'s study of yeast sporulation time series data (Raychaudhuri *et al.*, 2000) is an early example of application of PCA to microarray analysis. In this study, over 90% of the variance in the data was explained by the first two components of the PCA. The first principal component contained a strong steady-state signal. Projection scatter plots were used in an attempt to visualize previously identified gene groups, and to look for structures in the data that would indicate separation of genes into groups. No clear structures were visible that indicated any separation of genes in scatter plots. Holter *et al.*'s more recent SVD analysis of yeast sporulation data (Holter *et al.*, 2000) made use of a different pre-processing scheme from that of Raychaudhuri *et al*. The crucial difference is that the rows and columns of *X* in Holter *et al.*'s study were iteratively centered and normalized. In Holter *et al.*'s analysis, the first two eigengenes were found to account for over 60% of the variance for yeast sporulation data. The first two eigengenes were significantly different from those of Raychaudhuri *et al*., with no steady-state signal, and, most notably, structure indicating separation of gene groups was visible in the data. Below we discuss the discrepancy between these analyses of yeast sporulation data.

## 3. DISCUSSION

Selection of an appropriate pre-processing method is critical, and comparisons of results using different methods must always take the pre-processing into account. By inspecting the SVD of data, one can potentially evaluate different pre-processing choices by gaining insight into, *e.g.*, separability in scatter plots. The utility of SVD itself, however, depends on the choice of pre-processing, as the apparent discrepancy between the sporulation analyses described in section 2.3 illustrates. While structure was revealed in yeast sporulation data using the SVD on centered, normalized data (Holter *et al.*, 2000), structure was not visible using SVD on the original data (Raychaudhuri *et al.*, 2000), where the first component accounted for the steady-state gene expression levels. There are no hard rules to be applied, but in general the decision of how to pre-process the data should be made based on the statistics of the data, what questions are being asked, and what methods are being used to reveal information about those questions. As an example, performing a centering of gene transcriptional responses for time series data is often sensible because we are typically more interested in how a gene's transcriptional response varies over time than we are in its steady-state expression level.

An important capability distinguishing SVD and related methods from other analysis methods is the ability to detect weak signals in the data. Even when the structure of the data does not allow separation of data points, causing clustering algorithms to fail, it may be possible to detect biologically meaningful patterns. In section 2.2 we have given an example of this phenomenon using synthetic data. As an example of practical use of this kind of SVD-based analysis, it may be possible to detect whether the expression profile of a tissue culture changes in response to radiation dose, even when it is not possible to detect which specific genes change their expression in response to radiation dose.

SVD allows us to obtain the true dimensionality of our data, which is the rank *r* of matrix *X*. As the number of genes *m* is generally (at least presently) greater than the number of assays *n*, the matrix $V^T$ generally yields a representation of the assay expression profiles using a reduced number of variables. When $r < n$, the matrix *U* yields a representation of the gene transcriptional responses using a reduced number of variables. Although this property of the SVD is commonly referred to as dimensionality reduction, we note that any reconstruction of the original data requires generation of an $m \times n$ matrix, and thus requires a mapping that involves all of the original dimensions. Given the noise present in real data, in practice the rank of matrix *X* will always be *n*, leading to no dimensionality reduction for the gene transcriptional responses. It may be possible to detect the "true" rank *r* by ignoring selected components, thereby reducing the number of variables required to represent the gene transcriptional responses. As discussed above, existing SVD-based methods for pre-processing based on this kind of feature selection must be used with caution.

Current thoughts about use of SVD/PCA for gene expression analysis often include application of SVD as pre-processing for clustering. Clustering algorithms can be applied using, *e.g.*, the coordinates calculated for scatter plots instead of the original data points. Yeung and Ruzzo have characterized the effectiveness of gene clustering both with and without pre-processing using PCA (Yeung and Ruzzo, 2001). The pre-processing consisted of using PCA to select only the highest-variance principal components, thereby choosing a reduced number of variables for each gene's transcriptional response. The reduced variable sets were used as inputs to clustering algorithms. Better performance was observed without pre-processing for the tested algorithms and the data used, and the authors generally recommend against using PCA as a pre-processing step for clustering. The sole focus on gene clustering, however, in addition to the narrow scope of the tested algorithms and data, limit the implications of the results of this study. For example, when grouping assays is of interest, using $\{S\mathbf{a}'_j\}$ instead of $\{\mathbf{a}_j\}$ (see section 2; Figure 5.1) enables use of a significantly reduced number of variables (*r vs. m*) that account for *all* of the structure in the distribution of assays. Use of the reduced variable set for clustering must therefore result in not only decreased compute time, but also clusters of equal or higher quality. Thus the results in (Yeung and Ruzzo, 2001) for gene clustering do not apply to assay clustering.

In section 2.3 we discuss how, rather than separating into well-defined groups, cell-cycle genes tend to be more continuously distributed in SVD projections. For instance, when plotting the correlations of genes with the first two right singular vectors, cell-cycle genes appear to be relatively uniformly distributed about a ring. This structure suggests that, rather than using a classification method that groups genes according to their co-location in the neighborhood of a point (*e.g.*, *k*-means clustering), one should choose a classification method appropriate for dealing with ring-like distributions. Previous cell-cycle analyses therefore illustrate the fact that one important use of SVD is to aid in selection of appropriate classification methods by investigation of the dimensionality of the data.

In this chapter we have concentrated on conveying a general understanding of the application of SVD analysis to gene expression data. Here we briefly mention several specific SVD-based methods that have been published for use in gene expression analysis. For gene grouping, the *gene shaving* algorithm (Hastie *et al.*, 2000) and SVDMAN (Wall *et al.*, 2001) are available. An important feature to note about both gene shaving and SVDMAN is that each gene may be a member of more than one group. For evaluation of data, SVDMAN uses SVD-based interpolation of deleted data to detect sampling problems when the assays correspond to a sampling of an ordinal or continuous variable (*e.g.*, time series data). A program called SVDimpute (Troyanskaya *et al.*, 2001) implements an SVD-based algorithm for imputing missing values in gene expression data. Holter *et al.* have developed an SVD-based method for analysis of time series expression data (Holter *et al.*, 2001). The algorithm estimates a time translation matrix that describes evolution of the expression data in a linear model. Yeung *et al.* have also made use of SVD in a method for reverse engineering linearly coupled models of gene networks (Yeung *et al.*, 2002).

It is important to note that application of SVD and PCA to gene expression analysis is relatively recent, and that methods are currently evolving. Presently, gene expression analysis in general tends to consist of iterative applications of interactively performed analysis methods. The detailed path of any given analysis depends on what specific scientific questions are being addressed. As new inventions emerge, and further techniques and insights are obtained from other disciplines, we mark progress towards the goal of an integrated, theoretically sound approach to gene expression analysis.

## 4. FURTHER READING AND RESOURCES

The book (Jolliffe, 1986) is a fairly comprehensive reference on PCA (a new edition is meant to appear in summer of 2002); it gives interpretations of PCA and provides many example applications, with connections to and distinctions from other techniques such as correspondence analysis and factor analysis. For more details on the mathematics and computation of SVD, good references are (Golub and Van Loan, 1996), (Strang, 1998), (Berry, 1992), and (Jessup and Sorensen, 1994). SVDPACKC has been developed to compute the SVD algorithm (Berry *et al.*, 1993). Some web resources on SVD are found at the following URL's: http://www.cs.ut.ee/~toomas_l/linalg/; http://www.lapeth.ethz.ch/~david/diss/node10.html; and http://www.stanford.edu/class/cs205/notes/book/book.html. SVD is used in the solution of unconstrained linear least squares problems, matrix rank estimation, and canonical correlation analysis (Berry, 1992).

Applications of PCA and/or SVD to gene expression data have been published in (Alter *et al.*, 2000; Holter *et al.*, 2000; Holter *et al.*, 2001; Raychaudhuri *et al.*, 2000; Troyanskaya *et al.*, 2001; Yeung and Ruzzo, 2001; Yeung *et al.*, 2002). In addition, SVDMAN (Wall *et al.*, 2001) and gene shaving (Hastie *et al.*, 2000) are published SVD-based grouping algorithms; SVDMAN is free software available at http://home.lanl.gov/svdman. Knudsen illustrates some of the uses of PCA for visualization of gene expression data (Knudsen, 2002).

Everitt, Landau and Leese (Everitt *et al.*, 2001) present PCA as a special case of Projection Pursuit (Friedman and Tukey, 1974). Projection Pursuit, which in general attempts to find an "interesting projection" for the data, is also related to Independent Component Analysis (ICA) (Hyvärinen, 1999). ICA attempts to find a linear transformation (non-linear generalizations are possible) of the data so that the derived components are as statistically independent from each other as possible. Hyvärinen discusses ICA and how it relates to PCA and Projection Pursuit (Hyvärinen, 1999). Liebermeister has applied ICA to gene expression data (Liebermeister, 2002).

Other techniques that are related to PCA and SVD for visualization of data are Multidimensional Scaling (Borg and Groenen, 1997) and Self-Organizing Maps (SOM) (Kohonen, 2001). Both of these techniques use non-linear mappings of the data to find lower-dimensional representations. SOM's have been applied to gene expression data in (Tamayo *et al.*, 1999). There are also non-linear generalizations of PCA (Jolliffe, 1986; Scholkopf *et al.*, 1996).

## ACKNOWLEDGMENTS

We gratefully acknowledge Raphael Gottardo and Kevin Vixie for critically reading the manuscript. The writing of this chapter was performed within the auspices of the Department of Energy (DOE) under contract to the University of California, and was supported by Laboratory-Directed Research and Development at Los Alamos National Laboratory.

---

[1] For simplicity, we use the term *microarray* to refer to all varieties of global gene expression technologies.

[2] Complete understanding of the material in this chapter requires a basic understanding of linear algebra. We find mathematical definitions to be the only antidote to the many confusions that can arise in discussion of SVD and PCA.

[3] The *rank* of a matrix is the number of linearly independent rows or columns.

[4] The *covariance* between variables $x$ and $y$ is $C(x,y) = (N-1)^{-1}\Sigma_i(x_i-<x>)(y_i-<y>)$, where $N$ is the # of observations, and $i=1,…,N$. Elements of the *covariance matrix* for a set of variables $\{z^{(k)}\}$ are given by $c_{ij} = C(z^{(i)}, z^{(j)})$.

[5] A *centered* vector is one with zero mean value for the elements.

[6] Note that $(X^TX)_{ij} = \mathbf{a}_i \cdot \mathbf{a}_j$

[7] Note that $(XX^T)_{ij} = \mathbf{g}_i \cdot \mathbf{g}_j$

[8] A *normalized* vector is one with unit length.

[9] This notation is similar to that used in (Alter *et al.*, 2000), save that we use the term *eigenassay* instead of *eigenarray*.

[10] A *cluster center* is the average position of the points in a cluster.